\documentclass[a4paper]{PoS}

\usepackage{latexsym}
\usepackage{amsmath}
\usepackage{amsfonts,amssymb}
\usepackage{graphics,pgf}
\usepackage{rotating}
\usepackage{cite}

\newcommand{\beq}{\begin{equation}}
\newcommand{\eeq}{\end{equation}}
\newcommand{\be}{\begin{eqnarray}}
 \newcommand{\ee}{\end{eqnarray}}
\newcommand{\ov } {\over }

\def\vk{{\vec k}}

\def\ssmall{\fontsize{8pt}{8pt}\selectfont}

\title{High-energy physics and cosmological perturbations: observing
  new physics at large scales.}

\ShortTitle{High-energy physics and cosmological perturbations:
  observing new physics at large scales.}

\author{\speaker{Diego CHIALVA}%
         \thanks{The work of of DC is supported by the Belgian
           National ``Fond de la Recherche Scientifique"
           F.R.S.-F.N.R.S. with a contract ``charg\'e de
           recherche".}\\ 
        Universit\'e de Mons, Mons, Belgium\\
        E-mail: {diego.chialva@umons.ac.be}}

\abstract{Correlators of primordial perturbations could provide us
  with the signatures of physics at earlier times/higher momentum scales
  than inflation. The key-mechanisms are the interference and
  cumulation in time related to the interplay of negative- and
  positive-frequency components of fields and energy density generated by the
  high-momentum scale physics. 
  Here, we discuss which signatures are universal for
  such scenarios, and which ones instead would distinguish
  the specific cases (for example modified initial states for inflationary
  perturbations or modified dispersion relations). We also discuss
  the scale dependence of the correlators in presence of these
  signatures, especially for some scenarios, and how this could be
  interesting for observations.}

\FullConference{The European Physical Society Conference on High Energy Physics -EPS-HEP2013\\
		18-24 July 2013\\
		Stockholm, Sweden}

\begin{document}

\section{Introduction}

One of the main successes of the inflationary paradigm is the
prediction of ``seeds'' for structure formation, represented by
(quantum) fluctuations of the fields during inflation.
Several observations, present and soon-to-be, are sensitive to the
these fluctuations already at the perturbative level, opening
a window of opportunity to investigate
the physics during inflation. However, as we will discuss in 
this short communication, cosmological observables can be used as a diagnostic of the
physics at even higher momentum scales/ earlier times.

Research into these topics is of great importance nowadays, and not only for cosmology. Indeed, cosmological 
phenomena could be the only
reliable way to investigate very high-energy physics, providing the
arena to test theories and connect them with
observations.
This is even more significant now that the LHC has
confirmed the missing elements of the Standard Model of particle
physics, but has not given hints of new physics so far, while astrophysical
and cosmological observations present many challenges to our understanding.

In recent times there has been a large body of works on the topic
of high-energy signatures in inflationary perturbation theory
\cite{Chialva:2011hc, Chialva:2011iz, Ashoorioon:2011eg,
  Agullo:2010ws, SqueBisp, NonGausMoInSt, Transplanckian}. In this 
brief communication we will mainly deal with the general results after
\cite{Chialva:2011hc, Chialva:2011iz}, and discuss
universality and specificity of the
possible signatures.

\section{Cosmological observations and 
    quantum theory of perturbations.}


Cosmological observations agree well with a description of the
early universe in terms of a homogeneous isotropic
background with a Friedman-Robertson-Walker metric
(in conformal time)
 \beq
  ds^2 = a(\eta)^2 (-d\eta^2+ d\vec x^2), 
 \eeq 
and an inflationary sector driving an early accelerated expansion,
plus small perturbations. The latter ones are the
seeds for structure formation and are strongly constrained by
observations (initial energy density
contrast of only ${\delta\rho \ov \rho} \sim 10^{-5}$).

Observations and consistency of the theory require a quantum
description of the perturbations. In fact, due to the strong
redshift, classical perturbations would need an initial energy
density larger than the background one, thus invalidating
the perturbative approach well supported by data. In the quantum theory this
issue becomes just an indication that the initial state of the perturbations is the
vacuum or nearly so\footnote{Actually, quantum fluctuations have a non-zero vacuum energy
  density, which is in fact related to the
  cosmological constant issue. The usual attitude adopted when
  discussing inflationary perturbations is 
  to ignore this zero-point contribution, even
  though this is questionable in 
principle. 
}. 
Moreover, the inflationary redshift of physical scales is so
strong that the perturbations that we observe today had a
wavelength during inflation smaller than the Planck scale
(transPlanckian initial conditions \cite{Transplanckian}): 
this should require a
quantum treatment. Finally, the quantum theory 
explains the origin of the seeds themselves, as quantum
fluctuations always appear due to the uncertainty principle and do not
need to be postulated.

The cosmological
observables are gauge invariant expectation values of operators in
some initial state $|\Omega_{\text{in}}\rangle$. Using the interaction picture formalism:
 \beq \label{Correlator} 
  \langle\Omega_{\text{in}}| \widehat{\mathcal{O}}(\eta) |\Omega_{\text{in}}\rangle =
  \langle\Omega_{\text{in}}|\bar{T}\Bigl(e^{i\int_{\eta_{_\text{in}}}^{\eta}d\eta \widehat{\mathcal{H}}_I}\Bigr) \widehat{\mathcal{O}}_I(\eta)
  T\Bigl(e^{-i\int_{\eta_{_\text{in}}}^{\eta}d\eta \widehat{\mathcal{H}}_I}\Bigr)|\Omega_{\text{in}}\rangle\,,
 \eeq
where $(\bar{T})T$ is (anti)time-ordering, and, given the quadratic Hamiltonian of perturbations
 $\widehat{\mathcal{H}}_0(\eta)$,
 \beq \label{OperatorEq}
   i\partial_{\eta}\widehat{\mathcal{O}}_I(\eta) = 
      [\widehat{\mathcal{O}}_I(\eta), \widehat{\mathcal{H}}_0(\eta)].
 \eeq

\section{High-energy effects in cosmological perturbations.}

As we see from (\ref{Correlator}), observables are revealing of the physics
during inflation, in particular interactions,
background potential and
universe expansion (via $\widehat{\mathcal{H}}_I$ and $\widehat{\mathcal{H}}_0$).
However, they are also good diagnostic of earlier/higher momentum scale
physics. 

There are
two ways to study the effects of new physics. A top-down one where
the high-energy theory is modeled and the signatures are derived,
and a bottom-up one, which we will adopt, where the high-energy physics is parametrized and
its effects studied in terms of the parameters.
In particular we will focus on
1) the dependence of observables on the initial state
$|\Omega_{\text{in}}\rangle$, 2) the effects of
corrections to the kinetic  
terms at momentum scales higher than inflation. 
As we will see, the signatures of these features are not
irremediably washed out by the redshift during inflation.

We will deal with purely {\em adiabatic} perturbations, which are well-supported by
observations as possible leading seeds of structure. 
Remarkably, adiabatic perturbations are conserved when their physical
wavelength 
exceeds the Hubble scale, inverse of the Hubble rate 
$H \equiv {\partial_t a\ov a}$ ($\partial_t$ is the cosmic time derivative). 
 We will focus on scalar perturbations
described by the gauge invariant comoving curvature perturbation 
$\widehat{\mathcal{R}}(\vec x, \eta)$, see \cite{PertReviews} and
reference there for its precise definition,
and consider observables (\ref{Correlator}) given by correlators of this field.

The solution to the
field equation (\ref{OperatorEq}) for $\widehat{\mathcal{R}}$
(dropping the label ``$_I$'') is of the form
 \beq
   \widehat{\mathcal{R}}(\vec x, \eta) = \int {d^3k \ov (2\pi)^3}
    \bigl[{f_{k}(\eta) \ov z}\hat a_k + {f^*_{k}(\eta) \ov z}\hat a_k^\dagger\bigr].
 \eeq
where (primes stand for conformal time derivatives
and $\phi$ is the background inflaton)
 \beq  
  f_{\vec k}''(\eta) +(\omega({\vec k}, \eta)^2-{z'' \ov z})f_{\vec k}(\eta)=0 \,
  \qquad z \equiv {a\partial_t\phi \ov H}.
 \eeq 

The initial state $|\Omega_{\text{in}}\rangle$ in (\ref{OperatorEq}) and the form of the
dispersion relation $\omega({\vec k}, \eta)$ derived from
$\mathcal{H}_0$ dictate 
the high-momentum form of the mode
functions $f_{\vec k}$.
The standard choices are the adiabatic (also called Bunch-Davies) vacuum 
and the Lorentzian dispersion relation, defined as
 \beq \label{StandardChoices}
  \hat a_k(\eta_{\text{in}} \to -\infty)|\Omega_{BD} \rangle =0, 
  \qquad
  \omega(\vec k, \eta)^2 = k^2.
 \eeq

The choice of the adiabatic vacuum is motivated by its similarity 
with the Minkowski one at very small scales and by the demand for a certain
behaviour of the two-point functions (no antipodal poles).

However, the choices (\ref{StandardChoices}) can be questioned \cite{Transplanckian}. 
In fact, pre-inflation physics  
might have generated an excited initial state for inflationary
perturbations (still satisfying backreaction constraints). Furthermore,
from an effective theory viewpoint the states  
definition should not refer to physical scales way beyond the 
cutoff of the theory, as it is instead when $\eta_{\text{in}} \to -\infty$. 

Similarly, Lorentz invariance has been tested only up to certain scales. 
Moreover, there exist theoretical scenarios where that symmetry is broken at 
short scales (for example Ho\v{r}ava's theory). This motivates
investigating high-energy modifications to the dispersion relations
\cite{Transplanckian,Chialva:2011hc, Chialva:2011iz, Ashoorioon:2011eg}.
~~\hfill{} \qquad
\;\;We will now discuss general effects on observables from
modifications parametrizing new 
physics.

\section{Power spectrum}

Observations of the CMBR depict a nearly Gaussian statistics of
perturbations. An important observable is then the power
spectrum\footnote{Some authors define the spectrum rescaling the two-point function in our
  definition by $ {k^3 \ov 2\pi^2}$.}
after horizon exit ($\eta \to 0$) 
 \beq 
  P_{k} = \lim_{\eta \to 0} 
    \langle \mathcal{R}_{k}(\eta)\mathcal{R}_{k}(\eta)\rangle.
 \eeq
\paragraph{Standard scenario (labelled {\it s})} (adiabatic vacuum, Lorentzian dispersion, see
   (\ref{StandardChoices})). In a nearly de Sitter background, where
$\epsilon \equiv -{\dot H \ov H}  \ll1$, typical of slow-roll and
chaotic models, one finds
\vspace{-0.2cm}
  \beq \label{twopointstand}
  P_s(k) \underset{k \to 0}{\sim} 
   {H^2 \ov 4 M_{_\text{Planck}}^2 \epsilon k^3}\bigl({k \ov aH}\bigr)^{1-n_s},
  \eeq
\vspace{-0.2cm}
with $n_s \sim 0.96$ to comply with CMBR observations. $M_{_\text{Planck}}$ is the reduced Planck mass.
\paragraph{Modified initial state (labelled {\it mis}).} If at the initial time
$\eta_{\text{in}}$ some new physics above a momentum scale 
  $\Lambda \gg H$ sets different initial conditions than the adiabatic vacuum, the mode
    functions have the form
\vspace{-0.2cm}
 \beq
   f_{k}^{(\text{mis})}(\eta) = \alpha^{\text{mis}}(k, \Lambda, \eta_{\text{in}}) f^{\text{(s)}}_{k}(\eta) + \beta^{\text{mis}}(k, \Lambda, \eta_{\text{in}}) f^{\text{(s)}*}_{k}(\eta) \, ,
 \eeq
\vspace{-0.2cm}
in terms of those found in the standard scenario, here indicated as $f^{\text{(s)}}_{k}(\eta)$.
This leads to \cite{Transplanckian}
 \beq \label{twopointmodvac}
  P_{_\text{mis}}(k) \underset{k \to 0}{\sim} 
  {H^2 \ov 4 M_{_\text{Planck}}^2 \epsilon k^3}\bigl({k \ov aH}\bigr)^{1-n_s}
  \biggl(1+2 \, \text{Re}(\beta_k^{\text{mis}}e^{^\text{$i$Arg($\alpha_k^{\text{mis}}$)}}) \biggr)
 \eeq 
\paragraph{Modified dispersion relations (labelled {\it mdr}).} A dispersion relation with
corrections at a momentum scale $\Lambda \gg H$ can be written in general as 
\vspace{-0.2cm}
  \beq
    \omega(\eta, k) =   
      k \;  Q\left({H \ov \Lambda}k\eta\right) \, ,
    \qquad \text{with} \qquad Q(x \to 0) \to 1 \, .
   \eeq
It is found that the field correlators are quite different from the
standard scenario only if the adiabatic condition 
${\omega^\prime \ov \omega^2} \ll 1$ is 
violated at some early time \cite{Chialva:2011iz, Ashoorioon:2011eg}.
Then, even choosing initially the adiabatic vacuum, the field dynamics
lead to a spectrum  (see \cite{Chialva:2011iz} for the precise result for the mode functions) 
\vspace{-0.2cm}
 \beq
  P_{_\text{mdr}}(k) \underset{k \to 0}{\sim} 
  {H^2 \ov 4 M_{_\text{Planck}}^2\epsilon k^3}
  \biggl(1+2 \, \text{Re}(\beta^{^\text{mdr}}_{k}) \biggr) \, .
 \eeq
with $\beta^{^\text{mdr}}_k$ depending on $k$ and the interval of time
$\Delta$ where WKB is violated (constrained by backreaction).
Expanding for small $\Delta$, in full
generality $\beta^{^\text{mdr}}_k$ is
proportional to $\Delta$, with coefficient given by the parameter
signalling the violation of adiabaticity \cite{Transplanckian, Chialva:2011iz}. 

\subsection{Universal and specific signatures}

Both modified initial states and modified dispersion relations 
lead to a distinctive signature: a modulation
of the spectrum. The bottom-up approach we have been using
is not able to further constrain the aspect of the modulation. 
Detailed scenarios and models, with their specific $\beta$ coefficients, 
can be distinguished by the 
different pattern of modulation and their amplitudes.

From the point of view of the observer these signatures are interpreted
as the creation of particles\footnote{The ``particle" 
concept is not well-defined on time-evolving backgrounds, and neither
is energy in absence of a global timelike Killing vector,
but it is still useful as it relates easily to observations.} with an 
average number density of quanta
 $ 
    N_{\text{part}} \sim \int_k |\beta_k|^2
 $ 
leading to a pattern of interference due to ``positive'' and
   ``negative energy'' modes. 

\section{Non-Gaussianity}

The term non-Gaussianity indicates non-zero three- and higher-point correlators.
The leading perturbative one is the {\em bispectrum} 
$B(\vec k_1,\vec k_2,\vec k_3) \equiv 
    \langle
    \mathcal{R}(\vec k_1)\mathcal{R}(\vec k_2)\mathcal{R}(\vec k_3)\rangle|_{\eta \sim 0}$
traditionally parametrized
in terms of an amplitude $f_{NL}$ and a shape function with the momentum dependence. 
We write it as
  \begin{gather}
    B(\vec k_1,\vec k_2,\vec k_3, \eta) \equiv 
    (2\pi)^{\!\!^3} \delta(\sum_i \vec k_i)
  \Bigl(-{3 \ov 5}f_{NL}\Bigr)\Bigl({H^2 \ov 4\epsilon M_{_\text{Planck}}^2}\Bigr)^{\!\!^2}
  {4 \sum_i k_i^{-2} \ov k_t\prod_j 2k_j}F(\vec k_1, \vec k_2, \vec k_3) ;
  \quad 
  k_t\equiv  \sum_i k_i.
 \end{gather}

Clearly the bispectrum depends on the specific cubic couplings in the
theory. In the case of the Einstein-Hilbert theory the leading one in derivative expansion is 
$H_{I}^{(eh)}\, \text{{\tiny $=$}}\, \text{-}\!\!\int\!\! d^{3}\!\!x \, a^{3} ({\dot \phi \ov H})^{4} {H \ov M_{_\text{Planck}}^2}
    \, \mathcal{R}'^2 \vec\partial^{-2} \mathcal{R}'$. We take it as an example in the following, and consider again a quasi de Sitter background.
\paragraph{Standard scenario.} For single field slow roll/chaotic
models one finds \cite{Maldacena:2002vr}
\vspace{-0.2cm}
  \beq
   f_{NL} \approx -{\dot H \ov H^2} \ll 1, 
   \qquad
 \text{{\small $F_{\text{s}}(k_1, k_2, k_3, \eta)_{_\text{BD}}$}}  = 1
 \eeq
\paragraph{Modified initial state at a time
  $\eta_{\text{in}}$.}\footnote{If different modes have different
    initial times $\eta_{\text{in}}^{k_{1, 2, 3}}$, the relevant one
    for equations (\ref{BispMoInSt}), (\ref{ModiSqueBisp}) is the
    latest one.}
The bispectrum has on oscillating shape function, peaked on ``folded configurations"
  $ k_j = \sum_{h \neq j}k_h $, where the leading correction in
$\beta$-expansion reads
\cite{NonGausMoInSt}
 \beq \label{BispMoInSt}
  \text{{\small $\delta F_{_\text{mis}}(k_1, k_2, k_3)$}}  \!\sim  
   \text{{\small $ 
    -\!\!\sum_j\text{Re} \biggl[\beta_{k_j}^{\text{mis}^*}k_t{1-e^{i(\sum_{h \neq j}k_h-k_j)\eta_{\text{in}}} \ov \sum_{h \neq j}k_h-k_j}\biggr]
    \xrightarrow{k_j = \sum\limits_{h \neq j}k_h} \sim
    |\beta_{k_j}^{\text{mis}}| |k_t\eta_{\text{in}}| ; \qquad
    |k_t\eta_{\text{in}}| \gg 1 
    $}}
 \eeq
\paragraph{Modified dispersion relations.} Obviously a Lorentz-violating
   theory presents new cubic couplings, but, more importantly, the
   modified mode functions lead to a shape function with a more
   complicated oscillating piece, and peaked on those momenta
   configurations for which there is an $n$ and a time $\eta_*$ such that 
   $\partial_\eta^{m}\omega(k_j, \eta_*) = \sum_{h \neq j}\partial_\eta^{m}\omega(k_h, \eta_*)\; \forall m<n$ 
\cite{Chialva:2011iz}. 
For these configurations  
\vspace{-0.2cm}
 \beq 
  \delta F_{_\text{mdr}}(k_1, k_2, k_3) 
   \to  
   \sum_j \biggl({\Lambda \ov H}\biggr)^{1-{1 \ov n}}
   {1 \ov |O_{k_i, \eta_*}|^{{1\ov n}}}|\beta_{\vk_j}^{\text{mdr}^*}| \, ,
 \eeq
where $O_{k_i, \eta_*}$ is an order-1 function fully
specified in \cite{Chialva:2011iz}.

\subsection{Universal and specific signatures}

We can identify some generic types of signatures in the bispectrum
from modified initial states and dispersion relations. Some are  
common to the two scenarios:  an oscillating shape function, the possible presence of
enhancements (depending on the smallness of $\beta$), and greater
enhancements for interactions that scale with more powers of ${1 \ov a}$, which means a stronger  
sensitivity to higher derivative couplings.

Other features distinguish the two cases: different patterns of
oscillations and magnitude of ``enhancements", and non-Gaussianities
peaking for different configurations (in the case of modified
dispersion relations they are strictly tied to the specific pattern of
Lorentz violation \cite{Chialva:2011iz}). 

The physical interpretation of the signatures follows from three general points.
In the standard scenario the largest contribution to
   non-Gaussianities is at late times ($\sim$ horizon crossing),
but in modified high-energy physics scenarios
{\em particle creation} makes non-Gaussianities sizable
   also at early times. Furthermore, {\em interference} effects from
     ``negative/positive-energy'' components and
{\em cumulative} ones from
     time integration (consequence of time evolution) enhance these contributions.

\subsection{Squeezed limit and modified consistency relations}

Certain observations are sensitive to specific limits of the
bispectrum $B(\vec k_1,\vec k_2,\vec k_3)$.  For instance, the
power spectrum of peaks in the 
matter distributions at large scale
is related to the squeezed limit, where one of the wavenumber
is much smaller than the others: say, 
$k_1\text{{\ssmall $\ll$}}\,k_2\text{{\ssmall $\sim$}}\,k_3 \equiv k_S$. 

Such limit is also relevant theoretically, because various consistency
relations link the different correlators in such soft limits. The form
of the relations distinguishes types of models. However,
\cite{Agullo:2010ws, Chialva:2011hc} showed
that the relations do not depend only on the couplings and on the
background model, but also on the initial state of the perturbations
and on their dispersion relations.  

For example, for adiabatic perturbations one of these relations
connects the bispectrum to the spectrum. In the standard scenario the relation is determined by
powerful ``generic'' features:  non-Gaussianities peak at late times
(horizon crossing
), the ``squeezed"  $\mathcal{R}_{k_1}$ is superhorizon ($|k_1\eta| \ll 1$) at peak time and acts as
 background for the other perturbations shifting their exit-of-horizon times.
Thus, the limit is determined by the spectral index $n_s$:
\vspace{-0.2cm}
   \beq 
    \langle\mathcal{R}_{\vk_1}\mathcal{R}_{\vk_2}\mathcal{R}_{\vk_3}\rangle_{_\text{standard}}
    \underset{k_1 \ll k_S}{\simeq}
     (2\pi)^3 \delta(\sum_i\vk_i)\underbrace{(1-n_s)}_{O(\varepsilon)}  \underbrace{P_s(k_1)}_{\sim k_1^{-3}} P_s(k_S) \,
    \nonumber
   \eeq

\vspace{-0.2cm}
But in the cases of modified initial state ($\equiv$ {\it mis}) or modified dispersion
relations ($\equiv$ {\it mdr}) the new physics leads to new general features \cite{Chialva:2011hc}:
1) particle creation makes non-Gaussianities sizable 
  at earlier time $\eta_{ng} \text{{\small $\sim -$}}{\Lambda \ov Hk_{S}}$, 
enhanced by interference and 
cumulation in time,
2) for realistic $k_1$ (small but nonzero) it can be both
$|k_1\eta_{ng}| \gtrless 1$ depending on $\eta_{ng}$, 
   so $\mathcal{R}_{k_1}$ may be subhorizon
   at the time  of non-Gaussianities production. Thus, the squeezed limit
  need not be fully determined by $n_s$.

For example, for the Einstein-Hilbert cubic coupling $H_{I}^{(eh)}$,
see previous section, the standard result gets corrected by a
term (leading in $\beta$) \cite{Chialva:2011hc} 
\beq \label{ModiSqueBisp}
  \delta_\beta\langle\mathcal{R}_{\vk_1}\mathcal{R}_{\vk_2}\mathcal{R}_{\vk_3}\rangle
      \underset{k_1 \ll k_S}{\simeq} (2\pi)^3 \delta(\sum_i \vk_i) \, \,  
       4 \,\varepsilon \mathcal{B} \, P_s(k_1) P_s(k_S) \, ,
 \eeq
\vspace{-0.4cm}
 \be
  \mathcal{B}^{^\text{mis}}\!\!\!\!\!\! = \!\!\!\! \sum\limits_{j=2}^3\!\!
   \begin{cases} 
   \!\!\text{-} {k_S \ov  k_1} \,v_{_{\theta_j}}^{-1}
       \text{Re}\bigl[\beta^{^\text{mis*}}_{k_S} \bigr] \;
       \text{if $|k_1\eta_{_\text{in}} v_{_{\theta_j}}\!\!| \text{{\ssmall $\gg$}} 1$} \\
   \!\!\text{-} \,  k_S \eta_{_\text{in}} \,
      \text{Im}\bigl[\beta^{^\text{mis*}}_{k_S} \bigr] \;\;
       \text{if $|k_1\eta_{_\text{in}} v_{_{\theta_j}}\!\!| \text{{\ssmall $\ll$}} 1$}
  \end{cases}
  \;
  \mathcal{B}^{^\text{mdr}}\!\!\!\!\!\! = \!\!\!\! \sum\limits_{j=2}^3\!\!
  \begin{cases} \!\!\!\!
  \left. 
   \begin{aligned}
   & {\textstyle\!\,\text{-} \, {k_S \ov k_1} \,v_{_{\theta_j}}^{-1}
     \text{Re} \Bigl[\beta_{k_S}^{^\text{mdr*}}\Bigr]}
    & \!\!\!\! {\textstyle\text{{\small if $\vec k_1 \nparallel \vec k_j$}}} \\
    &
    \bigl({\Lambda \ov H}{k_1 \ov k_S}\bigr)^{\!\!\text{{\small${\kappa \ov \kappa+1}$}}} 
  {k_S \ov k_1}
   {\textstyle \text{Im} \Bigl[\beta_{k_S}^{^\text{mdr*}} \!\!\!\!\!\mathcal{O}(1)\Bigr] 
    }
    & \!\!\!\! {\textstyle \text{{\small if $\vec k_1 \parallel \vec k_j$}}}
   \end{aligned} \!\right]
    & \!\!\!\!\!\!\!\! \text{if ${\Lambda \ov H}{k_1 \ov k_S} \text{{\ssmall $\gg$}} 1$}
   \\[0.8cm]
   \; \; 4\, \varepsilon \,
   \, {\Lambda \ov H}
    \,\text{Im}\, \bigl[\beta_{k_S}^{^\text{mdr*}}\mathcal{O}(1) \bigr] 
    & \!\!\!\!\!\!\!\! \text{if ${\Lambda \ov H}{k_1 \ov k_S} \text{{\ssmall $\ll$}} 1$}   
  \end{cases} 
  \nonumber
 \ee
where $\kappa$ is the power of the leading correction in 
$\omega_{\text{phys}}(p) \sim p(1 + c_{_\kappa}\, \bigl({p \ov \Lambda}\bigr)^{\kappa} +\ldots)$,  
and $v_{_{\theta_j}} = (1+\cos(\theta_j))$  with
$\theta_j$ the angle between $\vec k_1$ and $\vec k_j$. 

\subsection{Universal and distinctive features in the squeezed limit}

It is possible to make an analysis for all couplings using the effective
theory for single-field inflation \cite{Chialva:2011hc}, to study universal
and distinctive signatures in the squeezed limit, say, 
$k_1 \ll k_{2, 3}\sim k_S$. 

In the {\em standard scenario} (adiabatic vacuum, Lorentz unbroken at
all scales) the bispectrum is local ($\sim k_1^{-3}$) in the squeezed
limit, and has
negligible amplitude $f_{_{NL}} \sim 1-n_s = O(\varepsilon)$
determined by the spectral index. 

With a {\em modified initial state}, the bispectrum has a non-local
form ($\sim k_1^{-n}, n = 4$), if 
$|k_1\eta_{\text{in}}v_{_{\theta_j}}| \text{{\ssmall $\gg$}} 1$
. 
In fact, one finds a local behaviour for a ``new squeezed limit'' 
$|k_1\eta_{\text{in}}v_{_{\theta_j}}| \text{{\ssmall $\ll$}} 1$
\cite{Chialva:2011hc}, but 
with a possibly enhanced amplitude (depending on the magnitude of $|\beta_{k_S}|$).

With a {\em modified dispersion
  relation}, we also find a new, stronger squeezed limit.
For ${\Lambda \ov H}{k_1 \ov k_S} \ll 1$ the leading contribution is local and the
   amplitude enhanced (depending on the magnitude of $\beta$) as $f_{NL} \sim \varepsilon {\Lambda \ov H}|\beta_k|$, while
higher derivative couplings, including Lorentz-breaking ones, are
suppressed by additional powers of ${k_1 \ov k_S}$.
For ${\Lambda \ov H}{k_1 \ov k_S} \gg 1$, the bispectrum is
  non-local, growing as $\sim k_1^{-4}$ for $k_1 \ll k_S$.
   Higher-derivative couplings can grow as $k_1^{-n}$ $n > 4$,
   but are suppressed by $\bigl({H \ov \Lambda}\bigr)^m$, $m \geq n$.

\section{Observations, comments and conclusions}

We have seen that primordial perturbations could provide us with a
window on physics at a very high-momentum scale.
The PLANCK collaboration has made a partial analysis \cite{Planck-data}
searching for
indications of modified initial states at the level of spectrum and
bispectrum, utilizing some templates. 
The results indicate that the templates  for 
the modified states scenarios give good fit to the 
spectrum data (even better than the standard scenario), but the
favoured part of the parameter space is fine tuned for
the present sensitivity of the data. The analysis of non-Gaussianity
(bispectrum) instead has not covered the whole (or large
representative samples of) parameter space, but only four specific
exemplary templates. The signal however has only a 1$\sigma$
significance. 

What about future Planck and after Planck? The search for
scale-dependent non-Gaussian signals hidden in the huge amount of data
could benefit from the type of bispectrum with modified initial states
or dispersion relations. In particular, modified dispersion relations
lead to a scale dependence tied to the pattern of Lorentz violations. 
Resonances could occur for example for light massive
fields. It is however necessary to reduce the parameter space, using
more top-down studies. Finally, LSS missions could tighten the
constraints on these high-energy modifications.

\end{document}